# Tuning the Magnetic Ground State by Charge Transfer Energy in SrCoO$_{2.5}$ via Strain Engineering

Sourav Chowdhury, Anupam Jana, R. J. Choudhary* and D. M. Phase

UGC DAE Consortium for Scientific Research, Indore-452001, India

SrCoO$_{2.5}$ (SCO) is a charge transfer insulator with $3d^6$ ground state configuration leading to antiferromagnetic nature. It is observed that substrate induced strain engineering modifies the ground state of SCO thin film with $3d^7\underline{L}$ ($\underline{L}$: O-2$p$ hole) configuration causing negative charge transfer energy ($\Delta$). The consequent strong hybridization between O-2$p$ and Co-3$d$ bands causes a hole in O-2$p$ band leading to hole mediated unconventional ferromagnetic ordering in SrCoO$_{2.5}$ thin film. This opens up a new avenue to tune the electronic structure vis a vis magnetic property via strain engineering.

Corresponding author: ram@csr.res.in

The exotic electronic and magnetic properties revealed by the transition metal oxides (TMOs) arise due to the competing forces encountered by transition metal (TM)-$3d$ electrons, such as electron-electron Coulombic interaction which favors localization of electrons, and TM $3d$-O $2p$ hybridization which facilitates the electrons delocalization. According to Zaanen-Sawatzky-Allen (ZSA) model [1], electronic properties of these materials hugely depend on the relative magnitude of the ligand to metal charge transfer energy $\Delta = E(d^{n+1}\underline{L}) - E(d^n)$ i.e. the energy required for $d^n \rightarrow d^{n+1}\underline{L}$ ($\underline{L}$: O-$2p$ hole), and the intra-atomic $d$-$d$ Coulomb repulsion energy $U_{dd}$ [1]. In the Mott-Hubbard regime, where $U_{dd} < \Delta$, the band gap corresponds to the charge fluctuations $d^n d^n = d^{n-1} d^{n+1}$ ($d$-$d$ type), with magnitude $\sim U_{dd}$. Whereas, in the charge transfer regime, where $\Delta < U_{dd}$, the band gap corresponds to the charge fluctuations $d^n d^n = d^n \underline{L} d^{n+1}$ ($p$-$d$ type), with magnitude $\sim \Delta$. If $\Delta$ is small compared to the O $2p$-TM $3d$ hybridization strength, the overlap between the $d^n$ and $d^{n+1}\underline{L}$ configurations becomes strong and the contribution of $d^{n+1}\underline{L}$ configuration to the ground state is considerable [2]. Accordingly, in highly covalent TMOs, the ground state domination is of $d^{n+1}\underline{L}$ configuration, consequently the charge fluctuations can be perceived as $d^{n+1}\underline{L} d^{n+1}\underline{L} = d^{n+1}\underline{L}^2 d^{n+1}$ ($p$-$p$ type). This type of material is regarded as the negative charge transfer material, wherein the band gap is neither a $d$-$d$ type, nor a $p$-$d$ type, rather it is a $p$-$p$ type [3-8]. Several interesting physical properties of various TMOs are owed to the negative charge transfer behavior such as, superconductivity in LaCuO$_3$ [9], metal-insulator transition in RNiO$_3$ (R: Rare earth elements) [10-12], diamagnetism in NaCuO$_2$ [3,13], ferromagnetism in SrCoO$_3$ [14-16].

Tuning $U_{dd}$ and $\Delta$ manifests striking modifications in the electrical and magnetic properties of TM based materials and demonstrates huge potential for applications [7,17-20]. $\Delta$ hugely depends upon the TM $3d$-O $2p$ hybridization and decreases with increase in hybridization. Thus, the TMOs with lower values of $\Delta$ and $3d^n$ ground state can be maneuvered to adopt a negative value of $\Delta$ and $3d^{n+1}\underline{L}$ ground state by tuning hybridization via strain in the system. Strontium cobaltates SrCoO$_{3-\delta}$ ($0 \leq \delta \leq 0.5$) belong to the strongly correlated electron systems, which have been hugely studied because of their well-endowed structural, electronic, and magnetic properties and reveal huge potential for catalyst, solid oxide fuel cell, and spintronic materials [21−24]. SrCoO$_3$ is known to show negative $\Delta$ and accordingly its ferromagnetic ground state is explained [15]. However, the electronic structure of SrCoO$_{2.5}$ (SCO) is rather unexplored, though its $\Delta$ value is believed to be small due to higher covalent character of Co$^{3+}$ based materials [25,26] and positive because of existing $Co^{3+}$-O-$Co^{3+}$ antiferromagnetic super-exchange interaction [15,27]. Thus, SCO is an ideal material to study the effect of modulation in $\Delta$ on the electronic and magnetic properties. In the present study using the combined valence band spectroscopy (VBS) and x-ray absorption spectroscopy (XAS) we show that $\Delta$ of SCO can be tuned via epitaxial strain in thin film form to a negative value, which interestingly leads to an unconventional ferromagnetic state. Tuning the position of SCO via strain in the ZSA diagram from positive to negative $\Delta$ regime and consequent changes in its magnetic ground state open up a new route for modifying the electronic and magnetic properties of various such materials to adopt novel properties with various technological potential applications.

To probe the electronic structure, we prepared a 30 nm thin film of SCO on (001) $SrTiO_3$ (STO) substrate. Details of the sample preparation and characterization methods are discussed in the supplementary material. X-ray diffraction (XRD) pattern confirms the single phase and epitaxial growth of the SCO film along *c*-direction (Fig. S1(a)). Due to the lattice mismatch with the STO substrate, the calculated pseudo-tetragonal out of plane lattice parameter of SCO film (3.940Å) is larger than that of bulk SCO (3.909Å) [21]. RHEED intensity oscillations during the deposition indicate a layer by layer growth of SCO film (Fig. S1(b)). The Co-2*p* core level x-ray photoelectron spectroscopy (XPS) reveals spin-orbit split $2p_{3/2}$ and $2p_{1/2}$ states at 781.3 and 796.8 eV binding energy positions (inset of Fig. 1(a)), indicating $Co^{3+}$ formal valence of SCO film [28,29]. Two other strong peaks at 786.3 and 803.3 eV correspond to the shake-up satellite of Co $2p_{3/2}$ and $2p_{1/2}$ states respectively [29].

Figure 1(a) shows the room temperature (RT) VBS of SCO film recorded at photon energy (PE) 52 eV. The observed VBS was fitted with the minimum number of Gaussian peaks, designated as features 1 to 8 using XPSPEAKFIT4.1 software, which adequately represent the major features of the spectrum. To distinguish the TM-3*d* state contribution to the VBS [30], we performed the resonance photoemission spectroscopy (RPES). For RPES, VBS is recorded at the PE values from 52 to 70 eV, swept through Co $3p \rightarrow 3d$ excitations. The variation of the spectral intensity for different features in the VBS with the incident PE is more clearly revealed in the constant initial state (CIS) plot (Fig. 1((b), (c)).

In RPES, the atomic O-2*p* photoionization cross-section decreases gradually with the PE, whereas, maximum photoemission intensity at a given PE occurs due to the interference between the direct Co-3*d* photoemission process: $Co3p^63d^6 + h\upsilon \rightarrow Co3p^63d^5 + e^-$ and intra-atomic photon induced excitation process: Co$3p^63d^6$ + $h\upsilon$ →Co [$3p^53d^7$]*, followed by the emission of a $3d$ electron through a super Coster Kronig decay, i.e. Co [$3p^53d^7$]* →Co$3p^63d^5$ + e$^-$ for Co-$3d$ derived state [31,32]. It is observed that the features 4 (5.0 eV) & 5 (5.9 eV) do not show any resonance (not shown here) confirming their O-$2p$ band nature. For La$_{1-x}$Sr$_x$CoO$_3$, the nonbonding O-$2p$ band occurs between 4.5 to 5.1 eV depending on the doping percentage [32,33], thus, feature 4 (5.0 eV) is ascribed as O-$2p$ nonbonding orbital. The features 1-3 (at 1.4, 2.1, and 3.1 eV respectively) and 6-8 (at 7.3, 9.9, and 12.3 eV respectively) show resonance phenomena, indicating their Co-$3d$ derived states (Fig. 1(b), (c)).

Generally, for $3d^nL$ final state, CIS spectra exhibits an anti-resonance dip, whereas for the $3d^{n-1}$ final state, CIS spectra exhibits a simple resonance enhancement without a considerable anti-resonance dip. To better understand this, the CIS spectra were fitted with the Fano line equation [34], as the line shape of the resonance,

$$I(w) = I_o(w)\frac{(q+E)2}{1+E^2} + I_{NR}(w)$$

where, I$_o$(w) is the $3d$ emission in the absence of the auto-ionizing transition, I$_{NR}$(w) is the noninterfering background contribution, $E = (h\upsilon - E_R)/\Gamma$: $E_R$ and $\Gamma$ being the energy and the width of the resonance respectively, and $q$ is the asymmetry parameter, determined by the magnitude and sign of the transition and interaction matrix elements. Generally, $q$ is lower for hybridized Co $3d$-O $2p$ structures and higher for pure Co-$3d$ structures, and is even higher for Co-$3d$ satellite states [35-37]. Features 1-3 show considerable dip in the CIS plot and fit well for lower $q$ values (0.37-0.29 respectively), indicating towards strongly hybridized O-$2p$ ($3d^nL$ final state) band character. Whereas, feature 6 reveals intense peak and $q$ =2.63 divulging its dominant Co-$3d$ ($3d^{n-1}$ final state) character. Features 7 & 8 exhibit even much higher $q$ values (4 & 4.2 respectively) confirming their Co-$3d$ satellite states (Fig. 1(b), (c)), which is also known to occur for cobaltates above 8 eV binding energy [26,32].

Co $L_{3,2}$-edge XAS of SCO film along with SCO bulk reference at RT are shown in figure 2(a). The Co-$2p$ absorption spectrum corresponds to the transition from the Co-$2p$ core level to the unoccupied Co-$3d$ states. In film a small hump (feature $S$) appears at the lower energy side of $L_3$ at PE ~778.5 eV, which arises due to the charge disproportionation (CD) (3-δ & 3+δ) present in the film, where δ depends on the TM $3d$-O $2p$ hybridization strength [38]. $L_{3,2}$ edge of bulk appears at the center of film's $L_{3,2}$ edges due to observed CD in film. Importantly, two feeble intense features $R1$ & $R2$ are also observed at the higher PE side of the $L_3$ and $L_2$ features. Its origin is discussed later in the manuscript. Such features are indicative of the negative Δ in the SCO film, as suggested by M. Abbate et al. [39] for $SrCoO_3$. Inset of figure 2(a) reveals O $K$-edge XAS of SCO film at RT. The O-1$s$ absorption spectrum corresponds to the transition from the O-1$s$ core level to the unoccupied O-2$p$ states hybridized with the different states of Co and Sr. The first broad feature at position 530.4 eV is the hybridized O-2$p$ band. We shall show later in the manuscript that this feature has dominant O-2$p$ character with the presence of O-2$p$ hole. The second feature at 534.9 eV is the O 2$p$-Sr 4$d$ band and a broad feature (537 to 546 eV) is attributed to the O 2$p$-Co 4$sp$ and O 2$p$-Co 4$s$ hybridized states [16,39,40].

In order to understand the band structure near the Fermi-level ($E_F$) of $SrCoO_{2.5}$ film, we combined the experimental valance band (VB) and the conduction band (CB) (O $K$-edge XAS) in a single frame (Fig. 2(b)). The attributions of different features in VB and CB are already discussed earlier in the manuscript. It is revealed that the upper most region in the VB and the lower most region in the CB are both hybridized O-2$p$ in nature suggesting a $p$-$p$ type of charge fluctuations. This indicates that the studied SCO film is in the negative Δ regime. Figure 2(c) shows the schematic band diagram of a negative Δ material (left) [3] along with the experimentally observed different bands of SCO film near the $E_F$ (right). From the observed diagram, the calculated value of Δ: the separation in energy between the center of O 2$p$-Co 3$d$ hybridized band i.e. $E_F$ and O-2$p$ band (feature 5) is -5.9 eV and also the value of $U_{dd}$: the separation in energy between $E_F$ (center of O 2$p$-Co 3$d$ hybridized band) and Co-3$d$ band (feature 6) is 7.3 eV (Fig. 2(c)). The observed negative value of Δ from the combined spectra is further confirmed by the CTM4XAS simulation [41] performed on Co $L$-edge (Fig. S2(a)). For simulation, the values of the parameters which provided the best match with the experimental spectrum are as follows: the crystal field energy =1.6 eV, $U_{2p3d} - U_{3d3d}$ =2.1 eV ($U_{2p3d}$: 2$p$ core hole-3$d$ electron Coulomb interaction), Δ =-6 eV, and ($pd\sigma$) hybridization =2.1 eV. It is also noted that the feeble features (*R1* & *R2*), vanish when positive value of Δ is considered. The Co $L$-edge XAS simulation for bulk SCO experimental spectra yields positive Δ ~3 eV, as expected and is shown in Fig. S2(b).

The strained films are known to exhibit unusual magnetic properties [20,42-45]. Due to the negative Δ of the strained SCO film and 3$d^7\underline{L}$ ground state, in contrast to the positive Δ of the bulk SCO with 3$d^6$ ground state, the magnetic property is envisaged to change from the antiferromagnetic nature in bulk. Magnetization (M) versus temperature (T) and magnetic field (H) at 5 K reveal ferromagnetic behavior of SCO film, with Curie temperature >360 K, saturation magnetization ($M_S$) value 1.8 $\mu_B$/f.u and coercivity 200 Oe (Fig. 3(a)). To understand the observed unusual ferromagnetic behavior, we write the ground state wave function for SCO as a linear combination of the possible states:

$$|\Psi> = \alpha|3d^6> + \beta|3d^7\underline{L}> + \gamma|3d^8\underline{L}^2> +….$$

where, for the negative $\Delta$ nature, $\beta$ must be greater than $\alpha$. Corresponding to the 100% $3d^7\underline{L}$, i.e. $\beta=1$, there is a presence of one hole in the O-$2p$ band per formula unit (SrCoO$_{2.5}$). This is equivalent to 1.6 and 2.4 number of holes in the CoO$_4$ tetrahedral and CoO$_6$ octahedral unit respectively. These created holes spin will align anti parallel to the Co-$3d$ electrons spin in order to conserve spin angular momentum during hopping. Interestingly, due to the nearest neighbor anti parallel alignment of O-$2p$ holes and Co-$3d$ electrons, Co-$3d$ spins will align parallel with each other, giving rise to ferromagnetic ordering (bottom inset of Fig. 3(a)) analogous to hole mediated ferromagnetic ordering in SrCoO$_3$ [15]. Practically, the weight of $3d^7\underline{L}$ configuration is much less than 100% in the SCO film, giving rise to two types of oxygen ions, one with the hole and other without hole. This will cause two different O-Co-O interactions. The interaction via an oxygen ion without a hole gives rise to antiferromagnetic super-exchange interaction, well-known for bulk SCO [27]. However, if there is a hole on the oxygen, the interaction is ferromagnetic because the hole facilitates the hybridization between the Co neighbors, as discussed above. Thus, the observed ferromagnetism in SCO film is an innate outcome of negative $\Delta$ of the film which could be due to the substrate induced strain. To further probe the effect of strain on $\Delta$ vis a vis magnetization of SCO film, we prepared another film of thickness 70 nm and studied their electronic and magnetic properties.

Interestingly, when thickness of the film is increased to 70 nm, $M_S$ drops to 0.6 $\mu_B$/f.u at 5 K (Fig. 3(a)). Due to the higher substrate induced strain in the thinner film than the thicker film (confirmed from XRD analysis), its in-plane Co-O-Co bond length is smaller causing greater charge transfer between O-$2p$ and Co-$3d$ bands. Therefore, the effective number of holes in the O-$2p$ band responsible for the parallel alignment of Co moments, is larger in the thinner film than the thicker film, leading to higher $M_S$ (1.8 $\mu_B$/f.u) for the thinner film.

Figure 3(b) shows the combined VBS and XAS spectra of 30 and 70 nm SCO films. It is observed that as strain relaxes partially with increase in thickness from 30 to 70 nm, $\Delta$ changes from -5.9 to -5.4 eV, revealed by the shift in O-$2p$ band (shown by the shaded area) by 0.5 eV. Furthermore, energy region 1 to 4 eV in VB is observed to show larger spectral weight for the thinner film signifying its larger O 2$p$-Co 3$d$ hybridization compared to the thicker film. In the CB, feature *A* reveals larger spectral weight for the thinner film (Fig. 3(b)), suggesting larger O-2$p$ hole density than the thicker film [20,45-48], which is also consistent with the variation in the $M_s$ and $\Delta$. Further, the onset of Co $L_{3,2}$-edge as well as its position is same in 30 & 70 nm films (Fig. 3(c)), suggesting Co$^{3+}$ state in both the films. Thus, the variation in feature *A* of O *K*-edge (Fig. 3(b)) with thickness affirms its dominant O-2$p$ nature.

The observation of ferromagnetism and the dependence of magnetization on the epitaxial strain appear to be an inherent outcome of the negative $\Delta$ character of the studied SCO films arising due to substrate induced strain. The claim is further substantiated from the Co *L*-edge spectra of different thickness SCO films (Fig. 3(c)), which divulge that while the spectral intensity of $L_3$ & $L_2$ features increases with thickness, that of *R1* & *R2* (attributed to the negative $\Delta$) decreases.

The variation in Co spectral features can be understood from the atomic multiplet calculations. The ground state of Co$^{3+}$ (3$d^6$) yields *L*=2 and *S*=2, therefore, *J* must be 4, i.e. a $^5D_4$ term symbol. For 2$p$ XAS with final state 2$p^5$3$d^7$, the allowed selection rule ($\Delta S$ =0, $\Delta L$ =0, ±1, $\Delta J$ =0, ±1) suggests $J^!$ to be 3, 4, or 5, which gives $L_{TOT}$ =2, 3, and 4; and $S_{TOT}$ =1 and 2. This provides one term with $J^!$ =5 ($^5F_5$) and two terms each with $J^!$ =3 and 4 ($^5F_3$, $^5D_3$, $^5F_4$, and $^5D_4$). Accordingly, five transitions are allowed forming Co $L_{3,2}$-edge (Fig. 3(c)). However, for negative $\Delta$ materials, the ground state of Co$^{3+}$ ion will also have the dominant contribution of 3$d^7\underline{L}$ besides 3$d^6$. Therefore, Hund's rule allows extra added level $^4F_{9/2}$ to the ground state and $^4G_{11/2}$, $^4G_{9/2}$, $^4G_{7/2}$, $^4F_{9/2}$, $^4F_{7/2}$, and $^4D_{7/2}$ to the excited state compared to the purely ionic Co$^{3+}$ configuration. Consequently, the extra transition from $^4F_{9/2}$ to $^4F_{9/2}$ & $^4F_{7/2}$ and from $^4F_{9/2}$ to $^4D_{7/2}$ are added at the higher energy side of the main $L_3$ and $L_2$ features respectively, resulting in features *R1* & *R2* (Fig. 3(c)). It is evident that with decrease in $\Delta$, the weight of $^4F_{9/2}$ configuration in the ground state will further increase and that of $^5D_4$ configuration will decrease causing higher spectral intensity of the *R1* & *R2* features as observed for the thinner film. This unambiguously establishes the role of thickness/ strain on the $\Delta$ and related spectral features.

It is apparent that the position in the ZSA diagram can be tuned via strain engineering in a compound for designing its electronic and magnetic properties. In SCO, $\Delta$ from positive for bulk is maneuvered to negative value through substrate induced epitaxial strain which triggers unconventional ferromagnetism. It is conceivable that $\Delta$ can be tuned through hydrostatic pressure also by doping with Ca or Ba such that the Co ions remain in the 3+ oxidation state. Here, it appears that the charge transfer gap collapses for negative Δ and the compound should show metallic behavior. However, as a result of strong *p-d* hybridization, the first ionization state is a $3d^7\underline{L}$ like discrete state split-off from the $3d^7\underline{L}$ continuum (Fig. 2(c)) [3,4]. Although, locally the $3d^7\underline{L}$ like split-off state is created but whether it remains stable in the periodic lattice i.e. the band gap remains open or collapses that only depends on the atomic arrangements [4,49]. In this study, the band gap remains open due to the strong intra-cluster (*pdσ*) hybridization (2.1 eV) and weak inter-cluster hybridization between the octahedral and the tetrahedral clusters, which drives the system to be in the negative Δ region, yet remaining in the insulating state.

In conclusion, the electronic structure of epitaxially strained $SrCoO_{2.5}$ film illustrates a *p-p* type of charge fluctuations with the negative Δ in contrast to its bulk counterpart. The consequent strong hybridization between O-2*p* and Co-3*d* bands causes a hole in O-2*p* band leading to hole mediated unconventional ferromagnetic behavior in $SrCoO_{2.5}$ film. Such modulation in electronic and magnetic properties via strain in SCO has huge potential for exploring other TMOs to adopt unusual properties. Indeed, heterojunction of such materials will lead to plethora of eccentric properties with huge potential for various applications.

## Acknowledgements:

The authors would like to thank Dr. Rajamani Raghunathan, Dr. V. R. Reddy, and Dr. Shailendra Kumar for valuable discussions. Authors also acknowledge Sharad Karwal, A. Wadikar, Rakesh Sah, and Manoj Kumar for their skilful technical assistance.

## References:

1. J. Zaanen, G. A. Sawatzky, and J. W. Allen, Phys. Rev. Lett. **55**, 418 (1985); J. Zaanen and G. A. Sawatzky, Can. J. Phys. **65**, 1262 (1987); J. Solid State Chem. **88**, 8 (1990).
2. M. Imada, A. Fujimori, and Y. Tokura, Rev. Mod. Phys. **70**, 1039 (1998).
3. T. Mizokawa, H. Namatame, A. Fujimori, K. Akeyama, H. Kondoh, H. Kuroda, and N. Kosugi, Phys. Rev. Lett. **67**, 12 (1991).
4. T. Mizokawa, A. Fujimori, H. Namatame, K. Akeyama, and N. Kosugi, Phys. Rev. B **49**, 11 (1994).
5. T. Mizokawa, D. I. Khomskii, and G. A. Sawatzky, Phys. Rev. B **61**, 11263 (2000).
6. D. I. Khomskii, J. Phys. Condens. Matter. **37**, 65 (1997).
7. S. Middey, J. Chakhalian, P. Mahadevan, J.W. Freeland, A.J. Millis, and D.D. Sarma, Annu. Rev. Mater. Res. **46,** 11.1 (2016).
8. J. Varignon, M. Bibes, and A. Zunger, Nat. Commun. **10**, 1658 (2019).
9. M. T. Czyzyk and G. A. Sawatzky, Phys. Rev. B **49**, 14211 (1994).
10. J. B. Torrance, P. Lacorre, A. I. Nazzal, E. J. Ansaldo, and Ch. Niedermayer, Phys. Rev. B **45**, 14 (1992).
11. J. L. García-Muñez, J. Rodriguez-Carvajal, and P. Lacorre, Europhys. Lett. **20**, 241 (1992); Phys. Rev. B **50**, 978 (1994).
12. S. Johnston, A. Mukherjee, I. Elfimov, M. Berciu, and G. A. Sawatzky, Phys. Rev. Lett. **112**, 106404 (2014).
13. K. Hestermann and P. Hoppe, Z. Anorg. Allg. Chemie. **367**, 261 (1969).
14. J. Kuneš, V. Křápek, N. Parragh, G. Sangiovanni, A. Toschi, and A. V. Kozhevnikov, Phys. Rev. Lett. **109**, 117206 (2012).
15. R. h. Potze, G. A. Sawatzky, and M. Abbate, Phys. Rev. B **51**, 17 (1995).
16. M. Abbate, G. Zampieri, J. Okamoto, A. Fujimori, S. Kawasaki, and M. Takano, Phys. Rev. B **65**, 165120 (2002).
17. S. B. Ogale, T. Venky Venkatesan, and Mark Blamir, *Fundamental Metal Oxides New Science and Novel Applications* (Wiley-VCH, Weinheim, 2013).
18. P. Yu, J.-S. Lee, S. Okamoto, M. D. Rossell, M. Huijben, C.-H. Yang, Q. He, J. X. Zhang, S.Y. Yang, M. J. Lee, Q. M. Ramasse, R. Erni, Y.-H. Chu, D. A. Arena, C.-C. Kao, L.W. Martin, and R. Ramesh, Phys. Rev. Lett. **105**, 027201 (2010).
19. P. A. Bhobe, A. Chainani, M. Taguchi, T. Takeuchi, R. Eguchi, M. Matsunami, K. Ishizaka, Y. Takata, M. Oura, Y. Senba, H. Ohashi, Y. Nishino, M. Yabashi, K. Tamasaku, T. Ishikawa, K. Takenaka, H. Takagi, and S. Shin, Phys. Rev. Lett. **104**, 236404 (2010).
20. H. Tanaka, Y. Takata, K. Horiba, M. Taguchi, A. Chainani, S. Shin, D. Miwa, K. Tamasaku, Y. Nishino, T. Ishikawa, E. Ikenaga, M. Awaji, A. Takeuchi, T. Kawai, and K. Kobayashi, Phys. Rev. B **73**, 094403 (2006).
21. V. R. Nallagatla, J. Jo, S. K. Acharya, M. Kim, and C. U. Jung, Sci. Rep. **9**, 1188 (2019).
22. H. Jeen, W. S. Choi, J. W. Freeland, H. Ohta, C. U. Jung, and H. N. Lee, Adv. Mater. **25**, 365 (2013).
23. H. Jeen, W. S. Choi, M. D. Biegalski, C. M. Folkman, I. C. Tung, D. D. Fong, J. W. Freeland, D. Shin, H. Ohta, M. F. Chisholm, and H. N. Lee, Nat. Mater. **12**, 1057 (2013).

24. N. Lu, P. Zhang, Q. Zhang, R. Qiao, Q. He, H. B. Li, Y. Wang, J. Guo, D. Zhang, Z. Duan, Z. Li, M. Wang, S. Yang, M. Yan, E. Arenholz, S. Zhou, W. Yang, L. Gu, C. W. Nan, J. Wu, Y. Tokura, and P. Yu, Nat. Lett. **546**, 125 (2017).
25. M. Abbate, R. h. Potze, G. A. Sawatzky, and A. Fujimori, Phys. Rev. B **49**, 7210 (1994).
26. T. Saitoh, T. Mizokawa, A. Fujimori, M. Abbate, Y. Takeda, and M. Takano, Phys. Rev. B **55**, 7 (1997).
27. T. Takeda and H. Watanabe, J. Phys. Soc. Jpn. **33**, 973 (1972).
28. F. Munakata, H. Takahashi, Y. Akimune, Y. Shichi, M. Tanimura, and Y. Inoue, R. Itti, and Y. Koyama, Phys. Rev. B **56**, 3 (1997).
29. Y. Wu, C. Dujardin, P. Granger, C. Tiseanu, S. Sandu, V. Kuncser, and V.I. Parvulescu, J. Phys. Chem. C **117**, 13989 (2013).
30. A. Jana, R. J. Choudhary, and D. M. Phase, Phys. Rev. B **98**, 075124 (2018).
31. C. Guillot, Y. Ballu, J. Paigne, J. I,ecante, K. P. Jain, P. Thiry, R. Pinchaux, Y. Petroff, and I. M. Falicov, Phys. Rev. Lett. **39**, 25 (1977).
32. S. Masuda, M. Aoki, Y. Harada, H. Hirohashi, Y. Watanabe, Y. Sakisaka, and H. Kato, Phys. Rev. Lett. **71**, 25 (1993).
33. A. Chainani, M. Mathew, and D. D. Sarma, Phys. Rev. B **46**, 16 (1992).
34. U. Fano, Phys. Rev. **124**, 6 (1961).
35. A. Fujimori, M. Saekl, N. Kimlzuka, M. Taniguchi, and S. Suga, Phys. Rev. B **34**, 10 (1986).
36. R. J. Lad' and V. E. Henrich, Phys. Rev. B **34**, 10 (1986).
37. R. J. Lad' and Victor E. Henrich, Phys. Rev. B **39**, 18 (1989).
38. M. Medarde, C. Dallera, M. Grioni, B. Delley, F. Vernay, J. Mesot, M. Sikora, J. A. Alonso, and M. J. Martínez-Lope, Phys. Rev. B **80**, 245105 (2009).
39. M. Abbate, L. Mogni, F. Prado, and A. Caneiro, Phys. Rev. B **71**, 195113 (2005).
40. S. Hu, Z. Yue, J. S. Lim, S. J. Callori, J. Bertinshaw, A. Ikeda-Ohno, T. Ohkochi, C.-H. Yang, X. Wang, C. Ulrich, and J. Seidel, Adv. Mater. Interfaces **2**, 1500012 (2015).
41. E. Stavitski and F. M.F. de Groot, Micron **1526**, 8 (2010).
42. J. S. White, M. Bator, Y. Hu, H. Luetkens, J. Stahn, S. Capelli, S. Das, M. Do¨beli, Th. Lippert, V. K. Malik, J. Martynczuk, A. Wokaun, M. Kenzelmann, Ch. Niedermayer, and C.W. Schneider, Phys. Rev. Lett. **111**, 037201 (2013).
43. K. Gupta and P. Mahadevan, Phys. Rev. B **79**, 020406(R) (2009).
44. A. Lupascu, J. P. Clancy, H. Gretarsson, Zixin Nie, J. Nichols, J. Terzic, G. Cao, S. S. A. Seo, Z. Islam, M. H. Upton, Jungho Kim, D. Casa, T. Gog, A. H. Said, Vamshi M. Katukuri, H. Stoll, L. Hozoi, J. van den Brink, and Young-June Kim, Phys. Rev. Lett. **112**, 147201 (2014).
45. H. Tanaka, Y. Takata, K. Horiba, M. Taguchi, A. Chainani, S. Shin, D. Miwa, K. Tamasaku, Y. Nishino, T. Ishikawa, E. Ikenaga, M. Awaji, A. Takeuchi, T. Kawai, and K. Kobayashi, Phys. Rev. B **80**, 081103(R) (2009).
46. Z. Hu, M. S. Golden, J. Fink, G. Kaindl, S. A. Warda, D. Reinen, P. Mahadevan, and D. D. Sarma, Phys. Rev. B **61**, 5 (2000).
47. F. M. F. de Groot, M. Gnom, J. C. Fuggle, J. Ghijsen G. A. Sawatzky, and H. Petersen, Phys. Rev. B **40**, 8 (1989).
48. P. Kuiper, G. Kruizinga, J. Ghijsen, M. Grioni, P. J. W. Weijs, F. M. F. de Groot, G. A. Sawatzky, H. Verweij L. F. Feiner, and H. Petersen, Phys. Rev. B **38**, 10 (1988).

49. T. Tsuyama, T. Matsuda, S. Chakraverty, J. Okamoto, E. Ikenaga, A. Tanaka, T. Mizokawa, H. Y. Hwang, Y. Tokura, and H. Wadati, Phys. Rev. B **91**, 115101 (2015).

**Figure captions:**

Figure 1. (a) Room temperature deconvoluted VBS of the SrCoO$_{2.5}$ film. Inset shows the Co-2$p$ XPS of SrCoO$_{2.5}$ film. (b) CIS plots of the features 1-3 and (c) CIS plots of the features 6-8; solid line shows the Fano line shape fitting.

Figure 2. (a) Room temperature Co $L$-edge XAS of SCO film along with bulk SCO reference. Inset shows the room temperature O $K$-edge XAS of SCO film. (b) Combined spectra of valence band and conduction band of SCO film showing a $p$-$p$ type of charge fluctuations. (c) The schematic band diagram of a negative Δ material and the experimentally observed different bands near the Fermi-level of SCO film. The opening and closing of the $p$-$p$ type band gap is shown in the figure for large and small value of $pd\sigma$ respectively.

Figure 3. (a) Magnetization (M) Vs field (H) curve of 30 & 70 nm SCO films at 5K. Top inset shows M-T curve in zero field cooled (ZFC) and field cooled (FC) cycles measured at H =1000 Oe. Bottom inset shows the schematics of Co-3$d$ parallel spin alignment via O-2$p$ hole. (b) Combined VBS and XAS spectra of 30 & 70 nm SCO films. (c) Effect of negative Δ on the atomic multiplet splitting (without considering the crystal field splitting), which leads to different spectral features in the Co $L$-edge XAS of 30 & 70 nm SCO films.

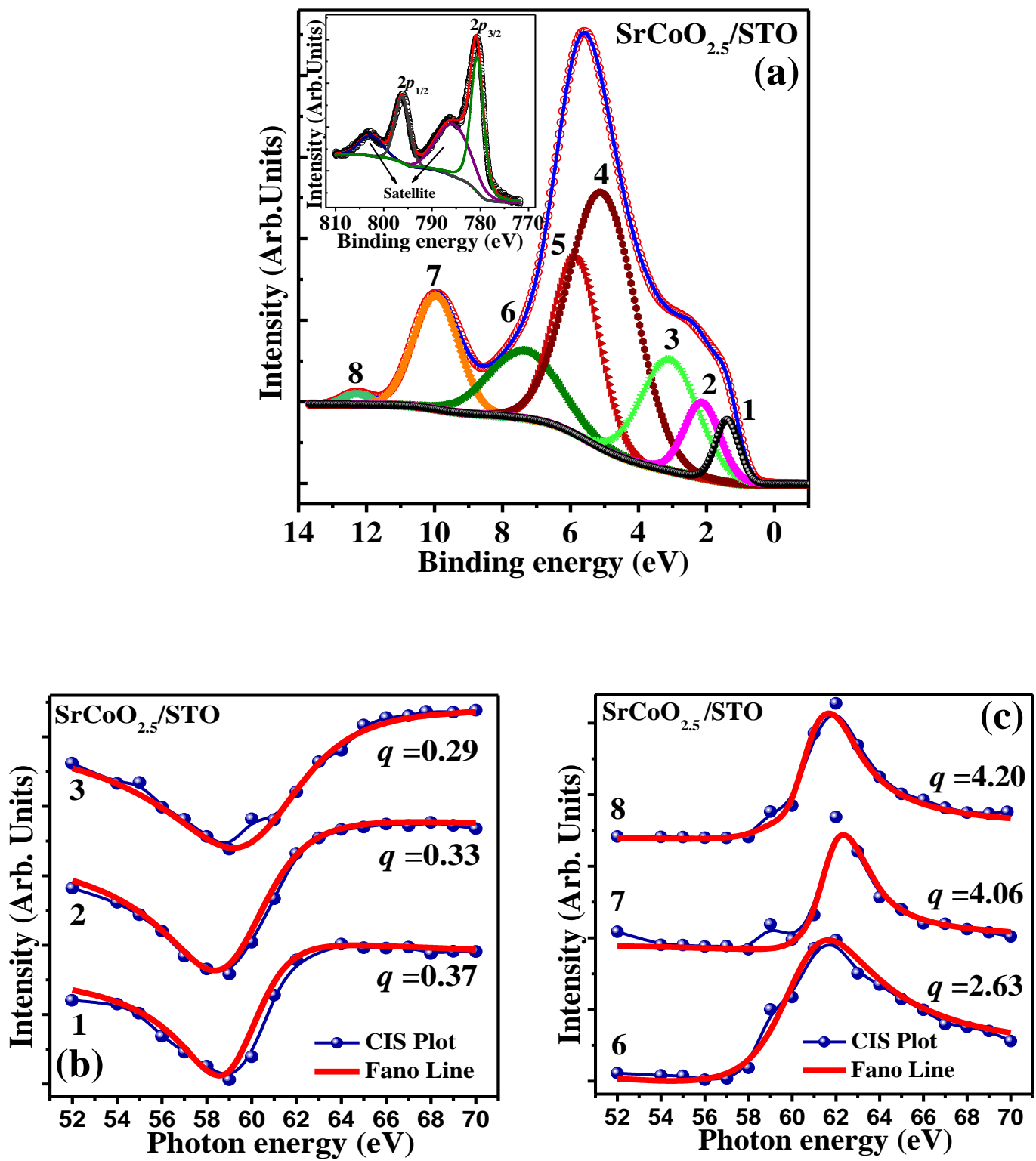

Fig. 1.

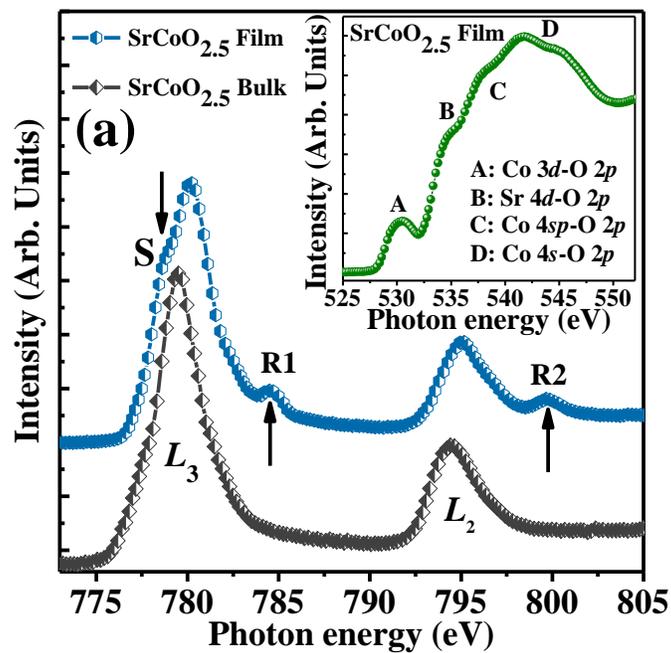
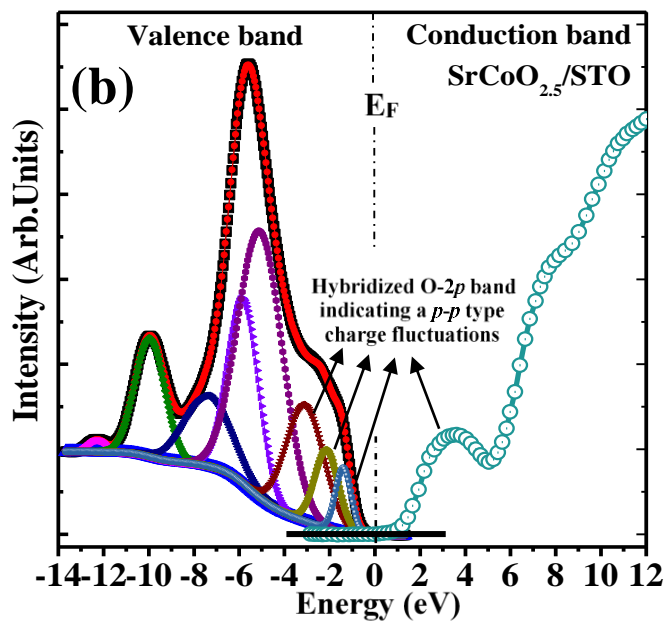
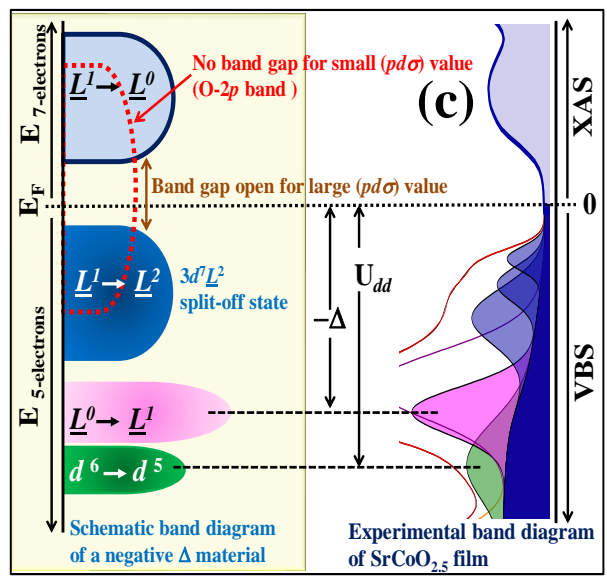

Fig. 2.

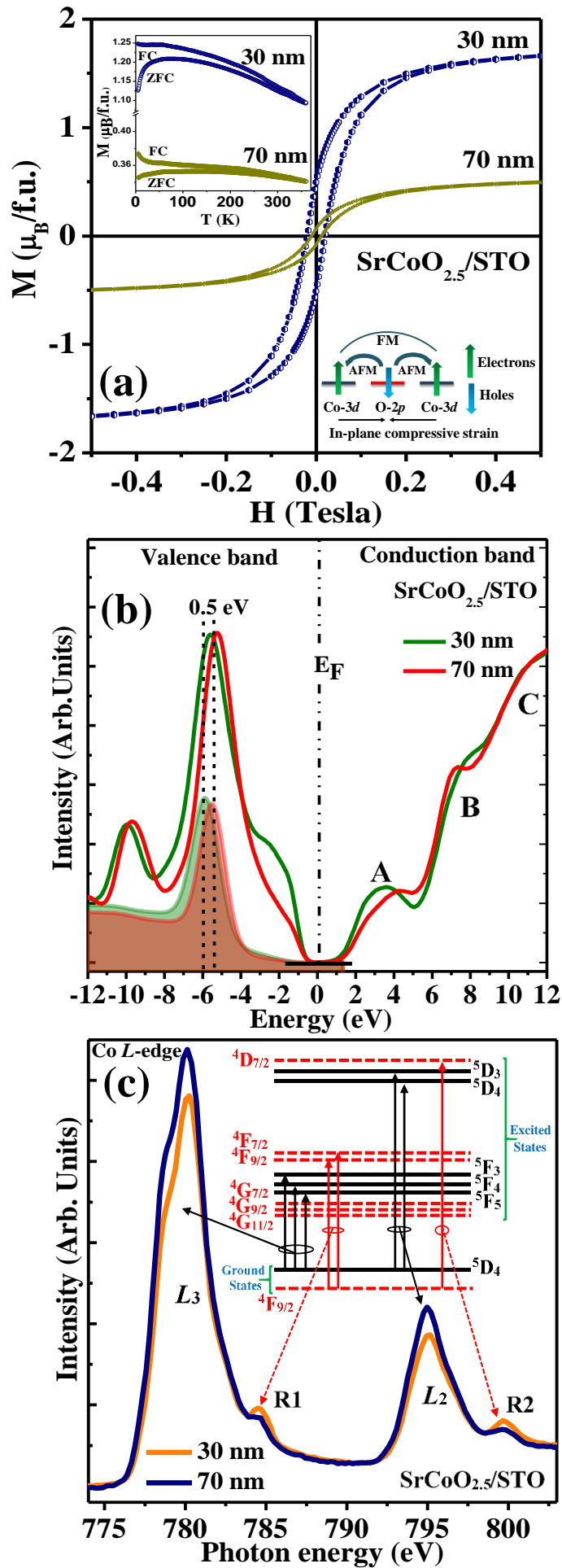

Fig. 3.